\documentclass{iopart}
\usepackage{iopams}  
\usepackage{graphicx}  % needed for figures
\usepackage{dcolumn}   % needed for some tables
\usepackage{bm}        % for math
\usepackage[english]{babel}

\newcommand{\be}{\begin{equation}}
\newcommand{\ee}{\end{equation}}
\newcommand{\bea}{\begin{eqnarray}}
\newcommand{\nn}{\nonumber}
\newcommand{\eea}{\end{eqnarray}}
\newcommand{\s}{\stackrel}

\begin{document}

\title{$f(R)$ gravity, torsion and non-metricity}

\author{Thomas P Sotiriou}

\address{Center for Fundamental Physics, University of Maryland, College Park, MD 20742-4111, USA}

\eads{\mailto{sotiriou@umd.edu}}

\begin{abstract}
For both $f(R)$ theories of gravity with an independent symmetric connection (no torsion), usually referred to as Palatini $f(R)$ gravity theories, and for $f(R)$ theories of gravity with torsion but no non-metricity, called $U4$ theories, it has been shown that the independent connection can actually be eliminated algebraically, as long as this connection does not couple to matter. Remarkably, the outcome in both case is the same theory, which is dynamically equivalent with an $\omega_0=-3/2$ Brans--Dicke theory. It is shown here that even for the most general case of an independent connection with both non-metricity and torsion one arrives at exactly the same theory as in the more restricted cases. This generalizes the previous results and explains why assuming that either the torsion or the the non-metricity vanishes ultimately leads to the same theory. It also demonstrates that $f(R)$ actions cannot support an independent connection which carries dynamical degrees of freedom, irrespectively of how general this connection is, at least as long as there is no connection-matter coupling.
\end{abstract}

\pacs{04.50.Kd, 04.20.Fy}
\maketitle

\section{Introduction}

$f(R)$ theories of gravity have received increased attention lately as modifications of General Relativity (GR) which can account for dark energy (see \cite{sfreview,phd,Sotiriou:2008ve,Nojiri:2006ri,Capozziello:2007ec} for reviews). The basic idea behind these theories is that the lagrangian is a general function of the Ricci scalar, instead of just being the Ricci scalar, as in GR. Two different variational principles have been applied to actions constructed from such lagrangians, as well as to the Einstein--Hilbert action: the more standard metric variation and an independent variation with respect to the metric and the connection, called Palatini variation. In the latter case, with which we will deal here, the connection is considered to be independent of the metric, at least to some extent. As can be found in textbooks (see {\em e.g.} \cite{grav}), if one assumes that the connection is symmetric and does not enter the matter action, Palatini variation and metric variation lead to the same field equations  for the Einstein--Hilbert action. However, this is not the case for more general actions.

Indeed, $f(R)$ theories with an independent, symmetric connection which does not couple to the matter, dubbed Palatini $f(R)$ theories of gravity, have been extensively studied \cite{palgen,fer} and are known not to be equivalent with the theory corresponding to the same action combined with simple metric variation. These theories have actually been shown to be equivalent with Brans--Dicke theory with Brans--Dicke parameter $\omega_0=-3/2$ \cite{flanagan,olmonewt,sot1}. This is a particular theory within the Brans--Dicke class in which the scalar does not carry any dynamics and can be algebraically eliminated in favour of the matter fields. As a matter of fact, in Palatini $f(R)$ gravity one can eliminate the independent connection algebraically even without any reference to the equivalent Brans--Dicke theory (see \cite{sfreview} for a discussion). The above imply that Palatini $f(R)$ gravity is a metric theory (according to Will's definition \cite{will}), which has also been shown explicitly \cite{koivisto}, and that the independent connection is really an auxiliary field \cite{sot2}.

Recently, a seemingly different theory was studied in \cite{capotor}, where the connection is allowed to be non-symmetric, and therefore allows torsion, but is forced to be metric, {\em i.e.}~it covariantly conserves the metric. It was shown there that this theory is also equivalent to $\omega_0=-3/2$ Brans--Dicke theory, and consequently to Palatini $f(R)$ gravity. Remarkably, this means that allowing the connection to have torsion and no non-metricity or non-metricity but no torsion leads to the same theory for $f(R)$ actions.

Even though this need not necessarily be the case, this might imply that even in the more general version of an $f(R)$ gravity theory with both torsion and non-metricity the connection will be an auxiliary field and one is ultimately led once more to the same theory. It will be shown here that this is indeed the case. The underlying reasons will also be thoroughly explained. In the process of our analysis we will also attempt to give a clear presentation of the differences between the aforementioned theories, as well as other similar theories in order to elucidate certain subtle issues.

Before going further, two clarifications are in order: Firstly, it has to be stressed once more that all theories mentioned above assume that the independent connection does not couple to the matter. Generalizations of Palatini $f(R)$ gravity in which the connection couples to the matter have been introduced in \cite{sotlib}. 
Secondly, the viability of models in the aforementioned theories certainly deserves a comment: It has been shown that Palatini $f(R)$ gravity models with infrared corrections with respect to GR are in conflict with the standard model of particle physics \cite{flanagan,padilla}. Additionally, the post-Newtonian metric in such models seems to have an algebraic dependence on the matter fields, and can lead to severe violation of solar system experiments \cite{olmonewt,sotnewt}. Finally, singularities have been shown to arise on the surface of well known spherically symmetric matter configuration \cite{Barausse:2007pn}, which render the theory at best incomplete and provide a very strong viability criterion. This criterion is almost independent of the functional form of the lagrangian, the only exception being lagrangians with corrections which become important only in the far ultraviolet (as in this case the singularities manifest at scales where non-classical effects take over) \cite{Olmo:2008pv} . All of the shortcomings just mentioned have essentially the same origin, which lies in the differential structure of the theory \cite{Barausse:2007pn}. Clearly, Any other theory which is equivalent to Palatini $f(R)$ gravity, such as $\omega_0=-3/2$ Brans--Dicke theory, or $f(R)$ gravity with torsion but no non-metricity will be plagued by the same problems. However, these issues will not concern us here, as our interest lies on the mathematical characteristics of these theories and not on their phenomenological implications. Our intention is to explore the features of theories with torsion and non-metricity and we focus on $f(R)$ actions just because they constitute well studied, simple toy models which could help in getting some insight into the matter.

\section{$f(R)$ actions and variational principles}

In this section we present a very brief review of the literature in order to clarify some subtleties and misconceptions. We follow the notation of Ref.~\cite{sotlib} and $\nabla_\mu$ denotes the covariant derivative of the affine connection $\Gamma^\lambda_{\phantom{a}\mu\nu}$ (not necessarily symmetric or metric), ${\cal R}^\mu_{\phantom{a}\nu\sigma\lambda}$ is the Riemann tensor defined in the usual way with this connection and ${\cal R}_{\mu\nu}\equiv {\cal R}^\sigma_{\phantom{a}\mu\sigma\nu}$ is the Ricci tensor (see \cite{sotlib} for ambiguities in the definition). The relation between the metric and the connection is characterized by the non-metricity, which measures the failure of the connection to covariantly conserve the metric, and the Cartan torsion tensor:
\be
\label{nonmet}
Q_{\mu\nu\lambda}\equiv-\nabla_\mu g_{\nu\lambda}, \qquad S_{\mu\nu}^{\phantom{ab}\lambda}\equiv \Gamma^{\lambda}_{\phantom{a}[\mu\nu]}.
\ee
%which measures the failure of the connection to covariantly conserve the metric and the Cartan torsion tensor
%\be
%\label{cartan}
%S_{\mu\nu}^{\phantom{ab}\lambda}\equiv \Gamma^{\lambda}_{\phantom{a}[\mu\nu]}.
%\ee
Square brackets (parenthesis) denote antisymmetrization (symmetrization). 

We are now ready to proceed with the variation of the action
\be
\label{action}
S=\frac{1}{2\kappa}\int d^4x \sqrt{-g} f({\cal R})+S_M,
\ee
where $\kappa=8\pi\, G$, $g$ is the determinant of the metric $g_{\mu\nu}$ and $S_M$ is the matter action.
There are two variational principles that one can apply to this action, the metric variation and the Palatini variation. The former  requires the {\em a priori} assumption that the connection is the Levi-Civita connection of the metric.
This implies that ${\cal R}=R$, where we use $R$ to denote the Ricci scalar associated with the Levi-Civita connection of the metric. Clearly, choosing metric variation does not allow any non-metricity or torsion right from the beginning.

The other alternative is to consider the connection independent of the metric (at least to some extend) and therefore, apply the so-called Palatini variation, which is an independent variation with respect to the metric and the connection. In this case there can be both torsion and non-metricity. However, before  proceeding further, one has to address the issue of the dependence of the matter action on the connection. Here there are two possible approaches. The first one  assumes that the matter action does not depend on the connection, {\em i.e.}~$S_M=S_M(g_{\mu\nu},\psi)$, where $\psi$ collectively denotes the matter fields. See \cite{sot2,sotlib} for the physical meaning of this assumption. This approach, usually dubbed the Palatini approach, has been extensively followed with the extra assumption that the connection is symmetric \cite{palgen,fer}. The second approach is to assume that the matter action does indeed depend on the connection, {\em i.e.}~$S_M=S_M(g_{\mu\nu},\Gamma^\lambda_{\phantom{a}\mu\nu},\psi)$. This approached, developed in \cite{sotlib}, is called the metric-affine approach. Note that the term metric-affine has sometimes been used to mean what we call here Palatini approach, see {\em e.g.} \cite{capotor}.

 Torsion and non-metricity in metric-affine $f(R)$ gravity has been studied extensively in \cite{sotlib} and we refer the reader there for details. For the rest of the paper we will focus on $f(R)$ theories of gravity without coupling between the matter and the independent connection as our goal is to understand the role of torsion and non-metricity in this class of theories. We wish to make no further assumption about the independent connections, apart from the fact that it is not coupled to the matter, in order to be as general as possible. 

Varying the action independently with respect to the metric and the connection gives the following set of field equations:
\bea
\label{field1}
&&f'({\cal R}) {\cal R}_{(\mu\nu)}-\frac{1}{2}f({\cal R})g_{\mu\nu}=\kappa T_{\mu\nu},\\
\label{field22}
&&\frac{1}{\sqrt{-g}}\bigg[-\nabla_\lambda\left(\sqrt{-g}f'({\cal R})g^{\mu\nu}\right)+\nabla_\sigma\left(\sqrt{-g}f'({\cal R})g^{\mu\sigma}\right){\delta^\nu}_\lambda\bigg]+{}\nn\\ & &+2f'({\cal R})\left(g^{\mu\nu}S^{\phantom{ab}\sigma}_{\lambda\sigma}-g^{\mu\rho}S^{\phantom{ab}\sigma}_{\rho\sigma}{\delta^\nu}_\lambda+g^{\mu\sigma}S^{\phantom{ab}\nu}_{\sigma\lambda}\right)=0,\eea
where a prime denotes differentiation with respect to the argument and, as usual,
$T_{\mu\nu}\equiv-2(-g)^{-1/2} \delta {S}_M/\delta g^{\mu\nu}$.
The right hand side of eq.~(\ref{field22}) vanishes thanks to our assumption that the matter action is independent of the connection. Details of the variation can be found in section 4.1 of Ref.~\cite{sotlib}.

By imposing further constraints on the connection one reduces to less general but also less complicated theories. The most typical example is Palatini $f(R)$ gravity without torsion. In this case one assumes {\em a priori} that the connection is symmetric. Then
%\be
%S_{\mu\nu}^{\phantom{ab}\lambda}=0,
%\ee
%and 
eq.~(\ref{field22}), after a series of manipulations (see {\em e.g.} \cite{sfreview}), can take the form
\bea
\label{gammagmn}
\Gamma^\lambda_{\phantom{a}\mu\nu}&=&\left\{^\lambda_{\phantom{a}\mu\nu}\right\}+\frac{1}{2f'}\Big[\partial_\mu f'\delta_{\nu}^\lambda+\partial_\nu f'\delta_{\mu}^\lambda-g^{\lambda\sigma}g_{\mu\nu}\partial_\sigma f'\Big].
\eea
Now notice that the trace of eq.~(\ref{field1}) is
\be
\label{trace}
f'({\cal R}){\cal R}-2f({\cal R})=\kappa T,
\ee
where $T=g^{\mu\nu}T_{\mu\nu}$. Since $f$ is a given function this is an algebraic equation in ${\cal R}$. We will not consider here the case where $f\propto {\cal R}^2$ and the left hand side is identically zero, which leads to a conformally invariant theory \cite{fer, sot1}, or the case where the equation has no root, as in this case there are also no solutions of the full field equations \cite{fer}. In all other cases, eq.~(\ref{trace}) can be used to express ${\cal R}$ as an algebraic function of $T$. However, this means that $f'$ in eq.~(\ref{gammagmn}) will also be an algebraic function of $T$. Therefore, the right hand side of this equation depends only on the metric and the matter fields. Then, $\Gamma^\lambda_{\phantom{a}\mu\nu}$ can actually be eliminated from eq.~(\ref{field1}) and the later can take the form
\bea
\label{eq:field}
\!\!\!R_{\mu\nu}- \frac{1}{2} R g_{\mu\nu}&= &\frac{\kappa}{f'}T_{\mu \nu}- \frac{1}{2} 
g_{\mu \nu} \left({\cal R} - \frac{f}{f'} \right) + \frac{1}{f'} \left(
			\s{g}{\nabla}_{\mu} \s{g}{\nabla}_\nu
			- g_{\mu \nu} \s{g}{\Box}
		\right) f'-\nn\\
& &- \frac{3}{2}\frac{1}{f'^2} \left[
			(\s{g}{\nabla}_{\mu}f')(\s{g}{\nabla}_{\nu}f')
			- \frac{1}{2}g_{\mu \nu} (\s{g}{\nabla} f')^2
		\right], 
\eea
where $\s{g}{\nabla}_{\mu}$ is the covariant derivative defined with the Levi-Civita connection,
$\s{g}{\Box}= g^{\mu\nu}\s{g}{\nabla} _{\nu}\s{g}{\nabla}_{\mu}$, and by virtue of eq.~(\ref{trace}) we know that $f$ and $f'$ are algebraic functions of $T$. We have successfully eliminated the independent connection completely.

Another example where the connection is not allowed to be the most general one is that when the non-metricity is assumed to vanish, {\em i.e.}~$Q_{\mu\nu\lambda}=0$. In this case, which has recently been considered in \cite{capotor}, eq.~(\ref{field22}) yields:
\be
S_{\mu\nu}^{\phantom{ab}\lambda}=\frac{\partial_\sigma f'}{4f'}\left(\delta^\sigma_\nu\delta^\lambda_\mu-\delta^\sigma_\mu\delta^\lambda_\nu\right), \qquad \Gamma^\lambda_{\phantom{a}\mu\nu}=\left\{^\lambda_{\phantom{a}\mu\nu}\right\}+S_{\mu\nu}^{\phantom{ab}\lambda},
\ee
and using eq.~(\ref{trace}) ${\cal R}$ can be expressed as an algebraic function of $T$ as before. Therefore, once more we can eliminate the independent connection completely and re-write eq.~(\ref{field1}) in the form of eq.~(\ref{eq:field}) \cite{capotor}. 

Remarkably, , in both simplified cases the independent connection can indeed be eliminated. Even more remarkably, assuming that there is no torsion and assuming that the is no non-metricity led to exactly the same field equations once the independent connection was eliminated. This hardly seems to be a coincidence. We will show next that even without any extra assumptions about torsion or non-metricity, the connection can still be eliminated and the field equation can take the form of eq.~(\ref{eq:field}).

\section{Eliminating the independent connection}

We return now to the most general case where no constraint have been applied to the connection. Following \cite{sand} we define the quantity
\be
\label{tG}
\tilde{\Gamma}^\lambda_{\phantom{a}\mu\nu}\equiv \Gamma^\lambda_{\phantom{a}\mu\nu}-\frac{2}{3}\delta^{\lambda}_\nu S_{\mu\sigma}^{\phantom{ab}\sigma}.
\ee
In terms of this quantity eq.~(\ref{field22}) can take the simple form
\be 
\partial_\lambda g_{\mu\nu}+\tilde{\Gamma}^\sigma_{\phantom{a}\lambda\nu}g_{\sigma\mu}+\tilde{\Gamma}^\sigma_{\phantom{a}\mu\lambda}g_{\sigma\nu}+\frac{\partial_\lambda f'}{f'}g_{\mu\nu}=0,
\ee
where we have taken into account the properties of differentiating densities and have utilized several algebraic manipulations (including contractions).
Exploiting the fact that the metric is  symmetric, this last equation can be written as
%\be
$\tilde{\nabla}_\lambda (f' g_{\mu\nu})=0$,
%\ee
where $\tilde{\nabla}_\mu$ is the covariant derivative associated with $\tilde{\Gamma}^\lambda_{\phantom{a}\mu\nu}$, or can be easily solved in terms of $\tilde{\Gamma}^\lambda_{\phantom{a}\mu\nu}$ to give
\bea
\label{gammagmn2}
\tilde{\Gamma}^\lambda_{\phantom{a}\mu\nu}&=&\left\{^\lambda_{\phantom{a}\mu\nu}\right\}+\frac{1}{2f'}\Big[\partial_\mu f'\delta_{\nu}^\lambda+\partial_\nu f'\delta_{\mu}^\lambda-g^{\lambda\sigma}g_{\mu\nu}\partial_\sigma f'\Big].
\eea
We stress once more that by virtue of eq.~(\ref{trace}) $f'$ can be considered an algebraic function of $T$, and, therefore, eq.~(\ref{gammagmn2}) gives $\tilde{\Gamma}^\lambda_{\phantom{a}\mu\nu}$ in terms of the metric and the matter fields only.

Let us now return to the definition of $\tilde{\Gamma}^\lambda_{\phantom{a}\mu\nu}$, eq.~(\ref{tG}). Using this equation one can easily show by performing contractions that 
\be
\tilde{\Gamma}^\lambda_{\phantom{a}[\mu\lambda]}=0.
\ee
This is enough to argue that determining $\tilde{\Gamma}^\lambda_{\phantom{a}\mu\nu}$ is not enough to fully determining $\Gamma^\lambda_{\phantom{a}\mu\nu}$, as eq.~(\ref{tG}) can never be solved to give the latter in terms of the former and its contractions. This should not come as a surprise. 

Let us consider the projective transformation
\be
\label{proj}
\Gamma^{\lambda}_{\phantom{a}\mu\nu}\rightarrow \Gamma^{\lambda}_{\phantom{a}\mu\nu}+{\delta^\lambda}_\mu\xi_\nu,
\ee
where $\xi_\nu$ is an arbitrary covariant vector field. One can easily show that the Ricci tensor and scalar will  correspondingly transform like
\be
\label{projRicci}
{\cal R}_{\mu\nu}\rightarrow {\cal R}_{\mu\nu}-2\partial_{[\mu}\xi_{\nu]}, \qquad {\cal R}\rightarrow {\cal R},
\ee
{\em i.e.}~${\cal R}$ is invariant under projective transformations. Hence any action built from a function of ${\cal R}$, such as action (\ref{action}), is projective invariant.
This implies exactly what we found above: that $\Gamma^{\lambda}_{\phantom{a}\mu\nu}$ can only be determined up to a projective transformation. 
 
 Indeed, determining $\tilde{\Gamma}^{\lambda}_{\phantom{a}\mu\nu}$ is enough for all practical purposes. Notice that only the symmetric part of the Ricci tensor of $\Gamma^{\lambda}_{\phantom{a}\mu\nu}$ enters eq.~(\ref{field1}). One can straightforwardly show though that
\bea
\label{tricci}
{\cal R}_{(\mu\nu)}&=&\partial_\lambda \Gamma^\lambda_{\phantom{a}(\mu\nu)}-\partial_{(\nu} \Gamma^\lambda_{\phantom{a}\mu)\lambda}+\Gamma^\lambda_{\phantom{a}\sigma\lambda}\Gamma^\sigma_{\phantom{a}(\mu\nu)}-\Gamma^\lambda_{\phantom{a}\sigma(\nu}\Gamma^{\sigma}_{\phantom{a}\mu)\lambda}\nn\\
&=&\partial_\lambda \tilde{\Gamma}^\lambda_{\phantom{a}(\mu\nu)}-\partial_{(\nu} \tilde{\Gamma}^\lambda_{\phantom{a}\mu)\lambda}+\tilde{\Gamma}^\lambda_{\phantom{a}\sigma\lambda}\tilde{\Gamma}^\sigma_{\phantom{a}(\mu\nu)}-\tilde{\Gamma}^\lambda_{\phantom{a}\sigma(\nu}\tilde{\Gamma}^{\sigma}_{\phantom{a}\mu)\lambda},
\eea
{\em i.e.}~the symmetric part of the Ricci tensor of $\Gamma^\lambda_{\phantom{a}\mu\nu}$ can be fully determined even if only $\tilde{\Gamma}^{\lambda}_{\phantom{a}\mu\nu}$ is known. This is already clearly exhibited in eq.~(\ref{projRicci}), as the symmetric part of the Ricci tensor is invariant under projective transformations.

Based on the above, the independent connection can indeed be eliminated from eq.~(\ref{field1}) using eqs.~(\ref{tricci}) and (\ref{gammagmn2}). The result is actually 
no different from eq.~(\ref{eq:field}) with ${\cal R}$, $f$ and $f'$ being algebraic functions of $T$ by virtue of eq.~(\ref{trace}). Therefore, the right hand side of eq.~(\ref{eq:field}) depends only on the matter fields and their derivatives and not on derivatives of the metric.

As an aside let as also point out the equivalence with Brans--Dicke theory. By making the field redefinition $\phi=f'$ and introducing the quantity
%\be
%\label{pot}
$V(\phi)=\phi{\cal R} - f$
%\ee
we can re-write eq.~(\ref{eq:field}) as
\bea
\label{bdf1}
G_{\mu \nu} &= &\frac{\kappa}{\phi}T_{\mu \nu}- \frac{1}{2} 
g_{\mu \nu}\frac{V(\phi)}{\phi} + \frac{1}{\phi} \left(
			\s{g}{\nabla}_{\mu} \s{g}{\nabla}_\nu
			- g_{\mu \nu} \s{g}{\Box}
		\right) \phi-\nn\\
& &- \frac{3}{2}\frac{1}{\phi^2} \left[
			(\s{g}{\nabla}_{\mu}\phi)(\s{g}{\nabla}_{\nu}\phi)
			- \frac{1}{2}g_{\mu \nu} (\s{g}{\nabla} \phi)^2
		\right].
\eea
On the other hand, eq.~(\ref{trace}) can be re-written as
%\be
$2V-\phi V'=\kappa T$,
%\ee
which combined with the trace of eq.~(\ref{bdf1})
yields
\be
\label{bdf2}
\s{g}{\Box}\phi=\frac{\phi}{3}(R-V')+\frac{1}{2\phi}\s{g}{\nabla}^{\mu}\phi\s{g}{\nabla}_{\mu}\phi.
\ee
Eqs.~(\ref{bdf1}) and (\ref{bdf2}) are the field equations can be derived by the action
\be
\label{bdaction}
S=\frac{1}{2\kappa}\int d^4x \sqrt{-g}\left(\phi R+\frac{3}{2\phi}\partial^\mu\phi\partial_\mu\phi-V(\phi)\right)+S_M
\ee
by varying with respect to $g_{\mu\nu}$ and $\phi$ respectively. Action (\ref{bdaction}) is the action of a Brans--Dicke theory with Brans-Dicke parameter $\omega_0=-3/2$ ($\omega_0$ is the numeric coefficient of the kinetic term in the action with a negative sign). This establishes the equivalence between $\omega_0=-3/2$ Brans--Dicke theory and the most general Palatini $f(R)$ gravity with both torsion and non-metricity. As already mentioned, the two restricted versions of Palatini $f(R)$ gravity, with vanishing torsion or vanishing non-metricity, have been known to be equivalent to $\omega_0=-3/2$ Brans--Dicke \cite{flanagan,olmonewt,sot1}.

This equivalence could have also been shown at the level of the action. 
Starting from action (\ref{action}) one can introduce a new field $\chi$ and write the dynamically 
equivalent action 
\be
\label{metactionH}
S=\frac{1}{ 2\kappa }\int d^4 x \sqrt{-g} 
\left[ f(\chi)+f'(\chi)({\cal R}-\chi)\right] 
+S_M.
\ee
 Variation with respect to $\chi$ leads to the equation
 \be
 \label{1600}
 f''( \chi )({\cal R}-\chi)=0.
 \ee
 Therefore,
  $\chi={\cal R}$ if
$f''(\chi)\neq 0$, which reproduces the 
action~(\ref{action}).
Redefining the field $\chi$ by $\phi=f'(\chi)$ and using $V$ as defined previously
  action (\ref{metactionH}) takes the form
\be
\label{metactionH2}
S_{met}=\frac{1}{ 2\kappa }\int d^4 x \sqrt{-g} \left[ \phi 
{\cal R}-V(\phi)\right] +S_M(g_{\mu\nu},\psi).
\ee
Since we could obtain $\tilde{\Gamma}^{\lambda}_{\phantom{a}\mu\nu}$  from the field equation derived by the initial action (\ref{action}) using purely algebraic manipulations, $\tilde{\Gamma}^{\lambda}_{\phantom{a}\mu\nu}$  is an auxiliary field and can be replaced in the action without introducing spurious solutions or affecting the dynamics. Using eqs.~(\ref{gammagmn2}) and (\ref{tricci}) and the definition of $\phi$ we can write
\be
{\cal R}=R+\frac{2}{3\phi^2}\partial^\mu \phi \partial_\mu \phi+\frac{3}{\phi}\s{g}{\Box}\phi
\ee
 and, therefore, ignoring a boundary term we can re-write action (\ref{metactionH2}) in the form of action (\ref{bdaction}).

\section{Discussion and conclusions}

In Palatini $f(R)$ gravity (symmetric independent connection) and $f(R)$ gravity with torsion but no non-metricity it was known that one can eliminate the connection in favour of the metric and the matter fields, as long as the connection does not enter the matter action. The outcome is in both case the same theory, which can also be written as an $\omega_0=-3/2$ Brans--Dicke theory. It has been shown here that even in the most general case where both torsion and non-metricity are allowed, the connection can still be algebraically eliminated, leading to the exact same theory as in the more restricted cases. 
Clearly, $f(R)$ actions do not carry enough dynamics to support an independent connection which carries dynamical degrees of freedom, as already discussed on \cite{Barausse:2007pn} for Palatini $f(R)$ gravity. 

It is important to realize that, even though in the theories considered here one starts with some connection different from the Levi-Civita connection of the metric, this connection turns out to be an auxiliary field. This implies that the geometry, at least as felt by matter, is {\em a priori} pseudo-Riemanian and that the theories are metric theories by construction (according to the definition of \cite{will}). Any geometry related to the independent connection is practically irrelevant for the matter. This of course might have been expected as the matter anyway couples only to the metric. In fact, one could say that the geometrical picture with torsion, non-metricity or both in such theories is to some extent simply misleading. Not only because it puts forward a redundant, complicated geometrical picture, but mostly, because in the independent connection representation of the theory the matter appears to be minimally coupled to the metric. However, as  is known already for Palatini $f(R)$ gravity, and holds for the more general case with both torsion and non-metricity as well since this ultimately leads to the same theory, the coupling between the matter and the metric is actually non-minimal \cite{flanagan, padilla, Barausse:2007pn}. This can be seen clearly once the independent connection has been algebraically eliminated, as the coupling of the latter to the metric introduces extra couplings between the matter and the metric, as well as self interactions for the matter fields.  See also Ref.~\cite{Sotiriou:2007zu} for an extended discussion on similar problems with gravity theories and their representations.

Before closing it is worth mentioning that, since the most general $f(R)$ theory with both torsion and non-metricity was found to lead to the same equation as Palatini $f(R)$ gravity and $\omega_0=-3/2$, it will certainly be plagued with the same serious viability issues as these theories, which were mentioned in the introduction.
As a final remark let us mention once more that the theories considered here assume that the connection does not couple to the matter. It would, therefore, be interesting to examine what happens in theories that do not include such an assumption, namely metric-affine $f(R)$ gravity \cite{sotlib}.

\section*{Acknowledgements}

This work was supported by the NSF Grant No. PHYS-0601800.

\end{document}